\documentclass[prd,nofootinbib,twocolumn,showpacs]{revtex4}
\usepackage{graphics,rotating,multirow,bm}

\def\etal{{\frenchspacing\it et al.}}

\def\eg{{\frenchspacing\it e.g.}}

\def\be{\begin{equation}}
\def\ee{\end{equation}}
\def\ba{\begin{eqnarray}}
\def\ea{\end{eqnarray}}

\begin{document}

\title{Tracking Dark Energy with the ISW effect: short and long-term predictions}

\author{Levon Pogosian$^1$, Pier Stefano Corasaniti$^2$,
Christian Stephan-Otto$^1$, Robert Crittenden$^3$, and Robert Nichol$^3$}

\affiliation{
$^1$ Institute of Cosmology, Department of Physics and Astronomy,
Tufts University, Medford, MA 02155, USA \\
$^2$ ISCAP, Columbia University, New York, NY 10027, USA \\
$^3$ ICG, University of Portsmouth, Portsmouth, PO1 2EG, UK}

\begin{abstract}

We present an analysis of the constraining power of future
measurements of the Integrated Sachs-Wolfe (ISW) effect on models of
the equation of state of dark energy as a function of redshift,
$w(z)$. To achieve this, we employ a new parameterization of $w$,
which utilizes the mean value of $w(z)$ ($\left< w \right>$) as
an explicit parameter.  This helps to separate the information
contained in the estimation of the distance to the last scattering
surface (from the CMB) from the information contained in the ISW
effect.  We then use Fisher analysis to forecast the expected
uncertainties in the measured parameters from future ISW
observations for two models of dark energy with very
different time evolution properties. 
For example, we demonstrate that the cross--correlation of 
Planck CMB data and LSST galaxy catalogs will provide competitive 
constraints on $w(z)$, compared to
a SNAP--like SNe project, for models of dark energy with a rapidly
changing equation of state (e.g. ``Kink'' models).
Our work confirms that, while SNe measurements are more suitable
for constraining variations in $w(z)$ at low redshift, the ISW
effect can provide important independent constraints on $w(z)$ at high
$z$.

\end{abstract}

\date{\today}

\maketitle

\section{Introduction}
\label{intro}

There is now a substantial amount of observational evidence that the
universe is dominated by a dark component which is causing
its expansion rate to accelerate.
The analysis of the
Cosmic Microwave Background (CMB) anisotropy power
spectra \cite{Spergel} combined
with results from large scale structure (LSS) surveys \cite{2dFSDSS}
strongly suggest that about $70\%$ of the energy in the universe
is in an exotic form of matter which we refer to as dark energy (DE).
Measurements of the luminosity
distance to Supernova Type Ia (SNIa) independently confirm these conclusions
by showing that high redshift supernovae are dimmer than
in a matter dominated universe \cite{RiessPerl}.

Further evidence, which has recently become available, is
the detection of the Integrated Sachs-Wolfe (ISW) effect using the
CMB/LSS cross-correlation \cite{xray,nolta,fosalba,scranton,afshordi03}.
This evidence is complementary in nature to the SNIa data.
Rather than probing the overall expansion of the universe,
it detects the slow down in the growth of density perturbations
that occurs when the universe ceases to be matter dominated.
CMB photons traveling to us from
the surface of last scattering
blueshift and redshift as they fall into and climb out of
gravitational potentials along their paths. During matter domination,
the large scale potentials remain constant in time, hence the
blueshift and the redshift exactly cancel each other out. However, any
deviation from a constant total equation of state results in a time variation of the
potentials and a net change in the photon energy. This is
observed as an additional CMB temperature anisotropy, which is the ISW effect \cite{Sachs}.

Detecting the ISW effect from measurements of CMB temperature anisotropy alone is not
feasible because the signal is hard to separate from the primordial anisotropy
from the last scattering surface.  To circumvent this, it was proposed to correlate
the CMB temperature maps
with the local distribution of matter \cite{turok96}.
The cross-correlation arises from CMB anisotropies produced
after the cosmic structures were formed. On large angular
scales the cross-correlation signal is dominated by the ISW effect,
while at small angles an additional contribution comes from
the Sunyaev-Zel`dovich effect \cite{Sun,Peiris}.

The cosmological constant $\Lambda$ is the
simplest model of dark energy that provides
a satisfactory fit to the existing data.
Despite the success of the concordance $\Lambda$ cold
dark matter scenario ($\Lambda$CDM), it is difficult to explain 
why the cosmological
constant value is so extremely small compared to the expectations of particle physics
without involving anthropic selection \cite{anthro}.
On the other hand, the observations are also consistent
with an evolving dark energy, such as Quintessence models where the dark energy
originates from a scalar field
\cite{wetterich,peebles,caldwell98}. Establishing whether the dark
energy is constant or evolving is one of the main challenges for
modern cosmology. In the Friedmann-Robertson-Walker (FRW) universe
the evolution of dark energy is completely determined
by its equation of state, which is defined as the ratio of pressure to
energy density: $w \equiv p/\rho$. For scalar field Quintessence, the
sound speed is unity and
$w$ determines the clustering properties of DE. Depending on the
model $w$ can be constant or change with time. Models with $w \ne
-1$ correspond to evolving DE, while $w = -1$ corresponds to
$\Lambda$.

The time dependence of
the dark energy equation has been constrained
by fitting various forms of $w(z)$
to the SNIa data, often in combination with CMB and the LSS measurements
\cite{Wang,Corasetal04,Hannestad04,Bruce04,Rapetti,jassal05}. However the data are not accurate enough to
distinguish between the cosmological constant and many forms of dynamical
dark energy. Moreover degeneracies between dark energy
parameters strongly limit the possibility to test whether $w$ is constant
or not.

As we shall describe in this paper, the
CMB/LSS correlation can provide another probe of the evolution of $w(z)$.
The sensitivity of the cross-correlation to the dark energy evolution
has previously been investigated in \cite{GPV04,Levon04}. In this paper
we study constraints expected from correlating surveys such as DES and LSST
with the CMB data. We introduce a novel way of parameterizing $w(z)$ that has the average 
$\left< w \right>$ (eq.~(\ref{def:aver}))
as an explicit parameter. This helps to separate information contained in the estimate of
the distance to last scattering from that in the ISW contribution.
We then proceed to show that the cross-correlation, in the absence of errors, 
is more sensitive to the details of the high redshift evolution of $w(z)$ than other available observables. 
This sensitivity, however, is off-set by large statistical errors due to a large primordial 
contribution to the CMB signal.
We forecast the expected uncertainties in DE parameters (using two different models)
extracted from CMB/LSS correlation that will be possible in the next five years (WMAP/DES)
and in the next ten years (Planck/LSST). We also calculate corresponding constraints from the
SNe data (SNLS in short term and SNAP in long term).
We find that in long term, cross-correlation can provide competitive constraints
on the evolution of $w(z)$.

The paper is organized as follows. In Section \ref{theory} we
discuss the dark energy modeling and the details of the calculations. Section \ref{sec:exp}
describes the observations considered in the paper.
In Section \ref{results} we discuss some advantages of including the cross correlation 
information and give the resulting parameter constraints.  Finally in Section
\ref{conc} we present our conclusions.

\section{Formalism and methods}
\label{theory}

\subsection {Dark energy model}\label{param}

Many models of dark energy have been proposed: cosmological constant ($\Lambda$),
Quintessence, k-essence, tangled defects, etc.
For simplicity, we will focus on quintessence based models where the
dark energy arises from a scalar field currently slow-rolling down a
potential, which is effectively inflation occurring at late times.
Perturbations in the dark energy are easily incorporated in such
models, and the sound speed is typically of order the speed of light.

The shape of the potential, which determines the evolution of the dark
energy, is generally a free function.  This makes it difficult to
predict the time dependence of the DE equation of state, so one is
potentially left with constraining a completely arbitrary function of
redshift.  The problem must be simplified by considering some
parameterization of the function $w(z)$.  Several parameterizations
have been suggested in a vast literature, but any resulting
constraints are very sensitive to the choice of parameterization.  For
instance it was shown in \cite{maor01} that fitting a constant $w$
will tend to favor lower negative values if the DE equation of state
is time dependent.  Even two-parameter Taylor expansions are too
simplistic, resulting in a strong dependence on the assumed priors and
generally biased against dynamical models \cite{Wang,Bruce04,Taxil04}.

One of the most popular two-parameter
formulae is the linear change in the scale factor $a=(1+z)^{-1}$
\cite{PolarskiLinder} given by, 
\be
w(a)=w_0+(a-1)\delta_w \ ,\label{lin}
\label{w:linder}
\ee

where $w_0=w_Q({\rm today})$ and $w_m = w_0 + \delta_w$ being the
value at some early time, such as the radiation-matter equality.  From
eq.~(\ref{lin}) it appears evident that with two parameters one can
fix the value of the equation of state today and at high redshift, but
the time evolution between the two extremes cannot be changed (i.e.
$dw/da={\rm const}.$).  As consequence of this, a linear expansion cannot
allow for rapid variation in the equation of state and particularly in
the case of Quintessence, for which $w_Q>-1$, this bias will tend to
favor small values of $\delta_w$.  Despite these potential problems,
eq.~(\ref{lin}) has become a common tool of testing the capability of
future experiments to constrain DE models (see \cite{UISLH}).  Due to
its wide use in the literature, we will consider this parameterization
here to help the comparison of our results with previous works. We
shall refer to it as Model~I.

In order to avoid the pitfalls introduced by the Taylor expansion we
also consider a phenomenological approach and model $w(z)$ according
to some minimal requirements, which are: 1) the form of $w(z)$ depends on a
minimal number of parameters which are of simple physical
interpretation, 2) the parameterization is not biased against any
particular time behavior and can account for both rapid (large) or
slow (small) variation of the equation of state, 3) the
parameterization reproduces the evolution of several proposed models
of DE such as Quintessence.  In this regard, a general formula 
constructed to follow these guidelines was proposed in
\cite{brucemeta,corased} and is usually referred to as the ``Kink'' model
\cite{Corasetal04}.  

\begin{figure}[tbp]
\centering
\scalebox{0.38}{\includegraphics{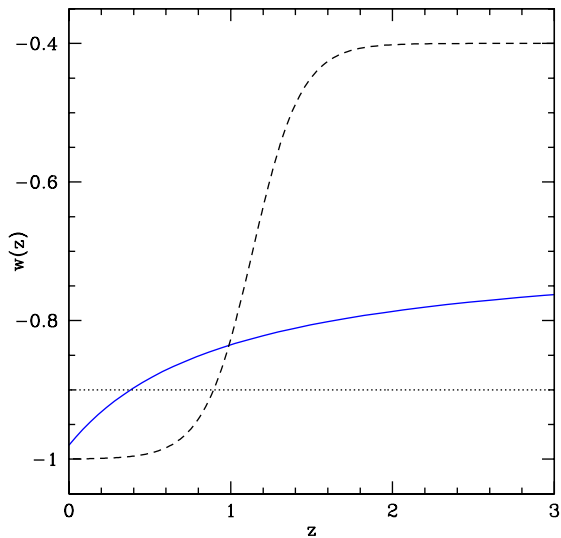}}
\caption{The equation of state as a function of redshift for the
two fiducial models considered in the paper. Model~I (blue solid line)
and Model~II (black dash line) have the same weighted average
$\left< w \right>=-0.9$ (the dotted line) defined in eq.~(\ref{def:aver}).}
\label{fig:eos}
\end{figure}

Within the class of the Kink models we consider a unique new
parameterization. We start with the functional form 
\be
w(z)= \frac{w_++w_-}{2} + \frac{w_+-w_-}{2} \ {\rm tanh}\left[{z-z_T
    \over \Delta_z}\right] \ ,
\label{def:tanh}
\ee
which describes a transition
from $w_+$ for $z\rightarrow +\infty$, to $w_-$ for
$z \rightarrow -\infty$,
with parameters $\Delta_z$ and $z_T$
describing the width and the central redshift
of the transition.
The novelty is in using the average equation of
state ($\left< w \right>$) as an explicit parameter, where $\left< w
\right>$ is defined as 
\be 
\label{def:aver} 
\left< w \right> \equiv
{\int_{a_{\rm LS}}^1 da \ w(a) \ \Omega_D(a) \over \int_{a_{\rm LS}}^1
  da \ \Omega_D(a)} \ .  
\ee 
Here $a$ is the scale factor ($a=1$
today), $a_{\rm LS}$ is the scale factor at the surface of last
scattering and $\Omega_D(a)$ is the ratio of the DE energy density to
the critical density.  It is well known \cite{huey99,dave02} that on
small scales the CMB temperature anisotropy spectra are insensitive
to the details of $w(z)$ except for its average value as defined in
eq.~(\ref{def:aver}).  
Specifying $\left< w \right>$ roughly fixes the
distance to the last scattering surface \cite{saini03}, which determines positions of
the acoustic peaks in the CMB temperature anisotropy spectrum.  The
CMB temperature data, combined with a prior knowledge of $H_0$ and under the
assumption that the universe is flat, is
capable of constraining the value of $\left< w \right>$ with a very
high accuracy.  
By explicitly specifying $\left< w \right>$ we
preserve the information that would be obtained from fitting a
constant $w$ to the CMB,
while still allowing for a varying $w(z)$. 
Hence, a direct control on the value $\left< w \right>$ allows us
to reduce the four dimensional parameter space to
a much smaller observationally allowed region, which corresponds to a narrow
range of possible values of $\left< w \right>$.

Note that for $\left< w \right>$ close
to $-1$, as favored by the data, most choices of $\Delta_w$ would lead to $w(z)$
changing between $-1$ and $-1+|\Delta_w|$, where $\Delta_w=w_--w_+$.
Also note that in a large
class of quintessence models either $w_+$ or $w_-$ is $-1$.
This is the case when the scalar field is initially in a slow-roll regime ($w_+=-1$) 
as in, \eg, the Doomsday model \cite{Linde86,GV00}. 
Similarly for tracker models, where the quintessence field starts tracking 
the background
component and evolves at late time settling to a minimum of its potential 
(i.~e. $w_{-}=-1$) \cite{Trackers}. Hence, fixing the value of either $w_-$ or $w_+$
to $-1$ is physically and observationally motivated and allows us to reduce the 
number of parameters in eq.~(\ref{def:tanh}) from four to three.

Thus, instead of using the four parameters of
eq.~(\ref{def:tanh}), $w_+$, $w_-$, $z_T$ and $\Delta_z$, we use three parameters:
$\left< w \right>$, $\Delta_w$ and $\Delta_z$.  Here, $z_T$ is chosen to
reproduce the desired value of $\left< w \right>$.
(Alternatively, we could explicitly specify $z_T$ and
use eq.~(\ref{def:aver}) to find the corresponding value of $\Delta_z$.)
Since we are primarily interested in determining whether $w(z)$ is varying or constant,
$\Delta_w$ is the main parameter of interest. The role of the third parameter, 
$z_T$ or $\Delta_z$, is only to allow for sufficient freedom in the ansatz.
In this paper, we will stay with the choice of
$\left< w \right>$, $\Delta_w$ and $\Delta_z$ and
refer to this parameterization as our Model~II.

Representative plots of $w(z)$ versus $z$ for Model~I and Model~II are shown in Fig.~\ref{fig:eos}.

\subsection{Temperature and density correlations}

The CMB temperature anisotropy due to the ISW effect
can be written as
\begin{equation}
{\delta T(\hat{\bf n}) \over \bar{T}}= \int_{\eta_m}^{\eta_0} d\eta
\ \left(\dot{\Phi}\left[(\eta_0-\eta) \hat{\bf n},\eta \right]
- \dot{\Psi}\left[(\eta_0-\eta) \hat{\bf n},\eta \right] \right),
\label{def:dt_isw}
\end{equation}
where $\hat{\bf n}$ is a direction on the sky, $\eta$ is the conformal
time, $\eta_m$ is some initial time far into the matter era, $\eta_0$
is the time today, $\Phi$ and $\Psi$ are the gravitational potentials
in the Newtonian gauge, and the dot denotes differentiation with
respect to $\eta$.

The ISW temperature anisotropy can be correlated with the distribution of
galaxies on the sky using,
\begin{equation}
\delta_g(\hat{\bf n}) = {{N}(\hat{\bf n})-\bar{N} \over \bar{N}}
\label{def:dg}
\end{equation}
where $\delta_g(\hat{\bf n})$ is the overdensity of galaxies in the
direction $\hat{\bf n}$, $N(\hat{\bf n})$ is the number of galaxies in
the pixel corresponding to the direction $\hat{\bf n}$, and $\bar{N}$
is the mean number of galaxies per pixel.  We assume here that we
sample galaxies in a fixed redshift region, characterized by a
normalized galaxy selection function, $W_g(z)$.  For simplicity, we
initially consider a single selection function, but below we consider
correlations arising between different redshift bins as well.

The galaxy number overdensity $\delta_g(\hat{\bf n})$ is assumed to be
tracing the cold dark matter (CDM) density contrast $\delta_c(\hat{\bf n})$ via
\begin{equation}
\delta_g(\hat{\bf n}) = b_g \delta_c(\hat{\bf n}),
\label{def:bias}
\end{equation}
where $b_g$ is the linear galaxy bias, which is possibly redshift dependent.
The CDM density contrast can be written as an integral over conformal time,
\begin{equation}
\delta_c(\hat{\bf n}) = \int_{\eta_m}^{\eta_0} d\eta \ {dz \over d\eta}
\ W_g(z(\eta)) \
{\delta_c}((\eta_0-\eta) \hat{\bf n},\eta) ,
\label{def:dc}
\end{equation}
where ${\delta_c}({\bf x},\eta)$ is the three-dimensional density contrast and is
directly related to the gravitational potentials.

We are interested in calculating the cross-correlation function
\begin{equation}
X(\theta) \equiv X(|\hat{\bf n}_1-\hat{\bf n}_2|) \equiv
\left\langle {\delta T^{\rm ISW} \over \bar{T}}(\hat{\bf n}_1) \delta_g(\hat{\bf n}_2)
\right\rangle \ ,
\label{def:xcor}
\end{equation}
where the angular brackets denote ensemble averaging and $\theta$ is
the angle between directions $\hat{\bf n}_1$ and $\hat{\bf n}_2$.
Calculations are often simplified when $X(\theta)$ is decomposed
into a Legendre series,
\begin{eqnarray}
X(\theta)=
\sum_{l=2}^{\infty} {2\ell+1 \over 4\pi} X_\ell P_\ell({\rm cos} \ \theta) \ .
\label{x_theta}
\end{eqnarray}
The coefficients $X_\ell$ can be evaluated using \cite{GPV04}:
\begin{equation}
X_\ell = 4\pi {9 \over 25}\int {dk \over k} \ \Delta_{\cal R} (k) \
\Delta^{\mathrm{ISW}}_\ell(k) \ M_\ell(k) \ ,
\label{eval:x_l}
\end{equation}
where $k$ is the wave-number, $\Delta_{\cal R}^2(k)$ is the primordial curvature
power spectrum, and functions $\Delta^{\mathrm{ISW}}_\ell(k)$ and $M_\ell(k)$ are defined as
\begin{eqnarray}
\Delta^{\mathrm{ISW}}_\ell(k) &=& \int_{\eta_k}^{\eta_0} d\eta \ j_\ell(k[\eta-\eta_0]) \nonumber
\\  &\times& (c_{\Phi \Psi} \dot{\phi}(k,\eta) - \dot{\psi}(k,\eta))\label{tisw}  \nonumber \\ 
M_\ell(k) &=& b_g c_{\delta \Psi} \int_{\eta_k}^{\eta_0} d\eta \
j_\ell(k[\eta-\eta_0])\nonumber \\  &\times&
\dot{z} W_g(z(\eta)) \delta (k,\eta) \label{mlk}\ .
\label{Delta_ISW}
\end{eqnarray}
In the above, the integration starts at an early time $\eta_k$,
when a given mode $k$ is well-outside the horizon, $j_l(\cdot )$ are
the spherical Bessel functions,
$\dot{\phi}(k,\eta)$, $\dot{\psi}(k,\eta)$ and
$\delta (k,\eta)$ are the evolution functions defined in \cite{GPV04}
along with the numerical coefficients $c_{\Phi \Psi}$ and $c_{\delta \Psi}$:
\begin{equation}
c_{\delta \Psi} \equiv {\delta \over \Psi} = -{3\over 2} \ , 
\ c_{\Phi \Psi} \equiv {\Phi \over \Psi} = -\left(1+{2\over 5} R_\nu \right)  \ ,
\label{numconst}
\end{equation} 
where $R_\nu \equiv \rho_\nu / (\rho_\gamma +\rho_\nu)$ and $\rho_\gamma$ and $\rho_\nu$ 
are the photon and relativistic neutrino densities.

Above we have considered the correlations between the CMB and a sample of galaxies 
defined by a single galaxy selection function.
Future surveys will enable us to separate galaxies into many redshift slices with 
photometric redshift slices (bins), and we can consider separate 
correlations with each of the slices, $X^i_\ell$.
We calculate these as above using the different selection functions, $W_g^i(z)$, 
each of which has a corresponding weighting function $M_\ell^i(k)$.  For each bin, 
we consider a 
different possible bias factor $b^i_g$, but as we discuss in Section \ref{sec:bias}, 
these can be determined sufficiently well by the observations of the 
galaxy-galaxy correlation functions.

The galaxy-galaxy correlations are evaluated in a similar way.  We have many 
correlation functions $\omega^{(i,j)}(\theta)$, corresponding to correlations 
between all possible redshift bins labeled by indices $(i,j)$.
Similarly to eq.~(\ref{eval:x_l}), the Legendre coefficients of these correlations 
are given by
\begin{equation}
\omega^{(i,j)}_\ell = 4\pi {9 \over 25}\int {dk \over k} \ \Delta^2_{\cal R} \
M^i_\ell(k) M^j_\ell(k) \ .
\label{def:m_ij}
\end{equation}

The full CMB auto-correlation $C^{TT}_\ell$ contains
the primary anisotropy from the last scattering surface as well as contributions from the
ISW, the SZ and other effects:
\begin{equation}
C^{TT}_\ell = C^{\rm LS}_\ell + C^{\rm ISW}_\ell + C^{\rm SZ}_\ell + ... \ .
\label{cl_sum}
\end{equation}
The uncorrelated anisotropies from the last scattering surface
dominate the noise in the cross correlation detection.  The ISW
contribution dominates the galaxy-CMB cross-correlation on large
scales (where the signal is strongest), while the SZ effect is
anti-correlated at WMAP frequencies and affects only the small angular scales \cite{fosalba}. 
In principle the SZ signal can be eliminated by smoothing
the CMB maps on scales smaller than the typical angular size of a
cluster (~$1^\circ$), therefore we will not include the SZ in our
analysis.

We evaluate the CMB/LSS correlation coefficients $X^{i}_\ell$,
as well as all other relevant spectra, using an appropriately modified
version of CMBFAST \cite{cmbfast}. For all models, we assume a flat
universe with adiabatic initial conditions for all particle species, including
the Quintessence.

\subsection{Fisher matrices}

We use the usual Fisher method \cite{fisher} for parameter estimation forecasts to
study the potential of the CMB/LSS cross-correlation for dark
energy constraints. We assume that the universe has evolved from
adiabatic initial conditions with a nearly scale-invariant spectrum
of density fluctuation, negligible tensor component and no massive neutrinos.
It is also assumed to be flat with a Quintessence field parameterized by $w(z)$.
In addition to the dark energy parameters $[w_0$,$\delta_w$] of
Model~I or [$\left< w \right>$,$\Delta_w$,$\Delta_z$] of Model~II, our parameter
space ${p_\alpha}$ includes the matter  fraction of total energy density $\Omega_M$,
the Hubble parameter $h$, the baryon density $\omega_b=\Omega_b h^2$,
the reionization optical depth $\tau$,
the scalar spectral index $n_s$ and the amplitude of the initial power
spectrum $A_s$. In total, we have $8$ parameters for Model~I and $9$
parameters for Model~II. In addition, for the SNe analysis, we 
account for $M$ -- the intrinsic supernova magnitude. Cross-correlation 
parameters also include the bias factors for each photometric bin, as
described in Section \ref{sec:bias}.

The cross-correlation Fisher matrix can be written as
\be
F^X_{\alpha,\beta}=f_{\rm sky} \sum_{\ell=\ell_{\rm min}}^{\ell_X} \sum_{i,j}
{\partial X^{i}_\ell \over \partial p_\alpha}
{\rm Cov}^{-1} (X_\ell^i X_\ell^j)
{\partial X^{j}_\ell \over \partial p_\beta}.
\label{fisher:x}
\ee
Here ${\rm Cov}^{-1} (X_\ell^i X_\ell^j)$, as well as the partial derivatives,
are evaluated at the fiducial values (see Sec.~\ref{sec:fid}). The sky fraction, $f_{\rm sky}$, 
is the smallest of the CMB and the galaxy survey sky coverage fractions.
We used $\ell_X=1000$ as the summation limit, although most of the
contribution comes from $\ell <500$. The smallest value of $\ell$ can be approximately
determined from $\ell_{\rm min}\approx \pi/(2f_{\rm sky})$.
The covariance matrix ${\rm Cov}_\ell$ is given by
\be
{\rm Cov}(X_\ell^i X_\ell^j) = \left(\tilde{C}^{TT}_\ell \tilde{\omega}^{(i,j)}_\ell 
+ X^{i}_\ell X^{j}_\ell \right)/(2\ell+1)  \ ,
\label{cov:x}
\ee where $\tilde{C}^{TT}_\ell$ and $\tilde{\omega}^{(i,j)}_\ell$ are
defined in eqns.~(\ref{cmb:noise}) and (\ref{galaxy:noise}) and
include the contributions from the noise.  Note that the contribution
to the $C^{TT}_\ell$ from the surface of last scattering is much
larger than that from the ISW effect, while $X^{j}_\ell$ is not
sensitive to the last scattering physics at all. This results in a
very large variance in $X^{j}_\ell$ that significantly impairs its
potential for parameter estimation.

For CMB measurements, the Fisher information matrix is given by
\be
F^{\mathrm{CMB}}_{\alpha,\beta}=f_{\rm sky}\sum_{\ell=2}^{\ell_\mathrm{CMB}}
\sum_{A,B}
{\partial \tilde{C}^{A}_\ell \over \partial p_\alpha}
\mathrm{Cov}^{-1}\big(\tilde{C}^{A}_\ell\tilde{C}^{B}_\ell\big)
{\partial \tilde{C}^{B}_\ell \over \partial p_\beta},
\label{fisher:cmb}
\ee
with the covariance matrices for $A,B=TT,EE,TE$ given explicitly in \cite{Zalda}. In the
above, $\ell_\mathrm{CMB}$ was taken to be $1200$ for WMAP and $2000$ for Planck, safely above
the maximum $\ell$ allowed by the angular resolution of each experiment.
The partial derivatives and the covariance matrix are evaluated at a fiducial choice
of parameters that is specified in the next subsection.

The quantity directly measured from SNe observations is their redshift-dependent
magnitude
\be
m(z) = M + 5 \log{d_L} + 25
\label{def:mz}
\ee
where $M$ is the intrinsic supernova magnitude and $d_L$ the luminosity distance (in Mpc).
The luminosity distance is defined as
\be \label{lumdis}
d_L(z)=(1+z) \int_0^z {d{z}' \over H({z}')} \ ,
\ee
where $H(z)$ is the Hubble parameter with a current value of $H_0$.
The information matrix for SNe observations is
\be
F^{\mathrm{SN}}_{\alpha,\beta}= \sum_{i}^{N}
{1\over \sigma(z_i)^2 }
{\partial m(z_i)\over \partial p_\alpha}
{\partial m(z_i)\over \partial p_\beta}.
\label{fisher:sne}
\ee
where the summation is over the redshift bins and $\sigma_m(z_i)$ is the value given by
eq.~(\ref{sigma_m}) at the midpoint of the $i$-th bin.

Given the Fisher matrix, the lower bound on
the uncertainty in determination of a parameter $p_i$ is given by
\be
\Delta p_i \ge C_{ii} = (F^{-1})_{ii} \ ,
\ee
where $C_{ij}$ is the error matrix.

One can forecast constraints from a combination of measurements
by adding the individual Fisher matrices. For example, cross-correlation
measurements can be combined with measurements of the CMB spectra by Planck
as
\be
F^{Planck+X}_{\alpha,\beta}=F^{Planck}_{\alpha,\beta}+
F^{X}_{\alpha,\beta} \ .
\ee

\subsection{The fiducial models}
\label{sec:fid}

\begin{figure}[tbp]
\centering
\scalebox{0.38}{\includegraphics{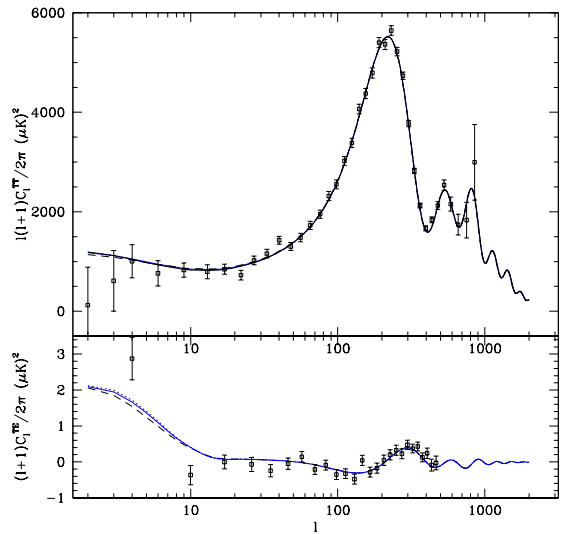}}
\caption{CMB temperature angular power spectrum (TT) and the
temperature-polarization cross-correlation (TE) for the three models
in Fig.~\ref{fig:eos} (using the same conventions for the three line types)
with the WMAP's first year data.}
\label{fig:cmb}
\end{figure}
\begin{figure}[tbp]
\centering
\scalebox{0.38}{\includegraphics{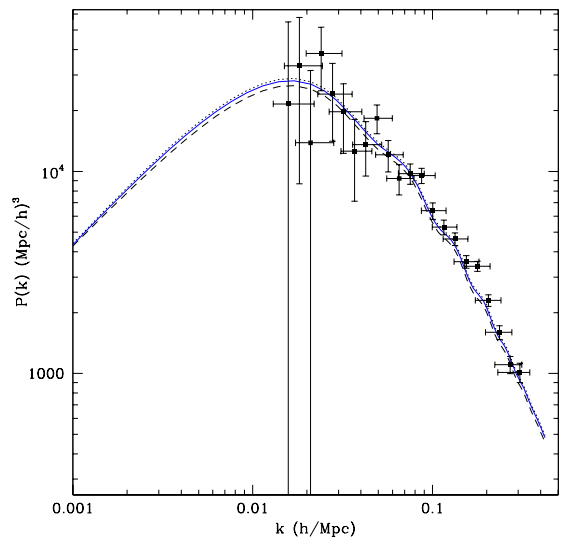}}
\caption{The linear matter power spectrum for the
three models in Fig.~\ref{fig:eos}. The SDSS data points \cite{sdss_pk}
are plotted to show current experimental error bars.}
\label{fig:pk}
\end{figure}
\begin{figure}[tbp]
\centering
\scalebox{0.38}{\includegraphics{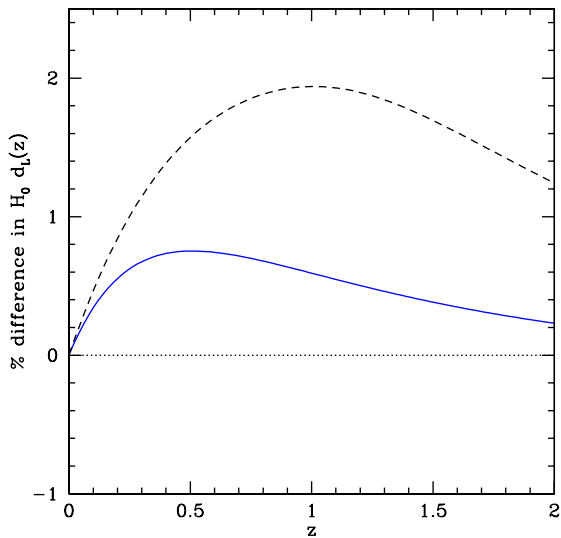}}
\caption{The percent difference in $H_0 d_L(z)$ between
the model with constant $w=-0.9$ (dotted line) and
Models I (solid blue) and Model~II (black dash). Current SNa
data can determine $H_0 d_L(z)$ with an accuracy of $5-10$\%.
Future data sets, such as supernovae from SNAP, can achieve an accuracy of $1-2$\%.}
\label{fig:lumin}
\end{figure}
Forecasts obtained using Fisher analysis are sensitive to the choice
of the fiducial model. We will consider the same fiducial models for
the two dark energy parameterizations discussed above and plotted in
Fig.~\ref{fig:eos}.  The fiducial parameters for Model~I were taken to
be $w_0=-0.98$ and $\delta_w=-0.29$.  For Model~II we chose $\left< w
\right>=-0.9$, $\Delta_w=-0.6$ and $\Delta_z=0.3$.  We take all other
cosmological parameters to be the same for both models. We have set
the Hubble parameter $h=0.69$, baryon density $\Omega_b h^2=0.024$,
total matter density $\Omega_M h^2=0.14$, spectral index $n_s=0.99$,
optical depth $\tau=0.166$ and the amplitude of scalar fluctuations
$A_s=0.86$ (as defined in \cite{wmap_verde}.)

All of current data is consistent with $w={\rm const}=-0.9$ at $1\sigma$ level
\cite{Spergel}. Both, Model~I and Model~II, have the same
weighted average $\left< w \right>=-0.9$. This assures that these
two models have nearly identical CMB spectra as shown in Fig.~\ref{fig:cmb}.

In Fig.~\ref{fig:pk} we compare the matter power spectra
predicted by the two models. Plotted are
the linear CDM spectra at $z=0$. The bias, non-linear effects and
the redshift space distortion would modify the three spectra in a similar way.
While one cannot compare these linear spectra to the data without accounting
for the above-mentioned corrections, we include the
SDSS data points \cite{sdss_pk} to illustrate the uncertainty in the current
determination of $P(k)$.

Models I and II are also consistent with current estimates of
the change in luminosity distance $d_L(z)$ obtained from the
SNIa measurements. 
In Fig.~\ref{fig:lumin} we plot the percent difference in $H_0 d_L$
between the model with constant $w=-0.9$ and Models I and Model~II.
The differences between the models are of order $1-2$\%, well below
the current level uncertainty and comparable to the projected
accuracy of SNAP.

Hence, present data cannot distinguish between $w={\rm const}=-0.9$ and our
Models I and II and both are perfectly consistent with all available observations.

\section{Experiments}
\label{sec:exp}

We are interested in forecasting the errors in dark energy
parameters that can be extracted from CMB/LSS correlation studies
and compare them to those expected from the luminosity distance measurements.
We make short- (less than 5 years) and long- (ten years) term predictions based on
the ongoing and planned CMB, galaxy and SNe observations.

For our short term predictions we assume the 4-year CMB temperature and polarization
data from WMAP \cite{wmap}, the expected galaxy counts from the Dark Energy 
Survey (DES) \cite{des},
and the SNe from the Nearby Supernova Factory (NSNF) \cite{nsfactory} and
the Supernovae Legacy Survey (SNLS) \cite{snls}. In addition, we impose a 
Gaussian prior on the value of $h$ corresponding to a $1\sigma$ uncertainty of
$\pm 0.08$ from the Hubble Space Telescope's Key Project (HSTKP) \cite{hconstraint}.

For the long term predictions we assume the 14 months CMB temperature and polarization
data by Planck satellite \cite{planck}, the galaxy catalogues by the 
Large Synoptic Survey Telescope (LSST) \cite{lsst}, and the SNe from
the Supernovae Acceleration Probe (SNAP) \cite{snap} complemented by those 
from the NSNF.

\subsection{Galaxy data: DES and LSST}
\begin{figure}[tbp]
\centering
\scalebox{0.38}{\includegraphics{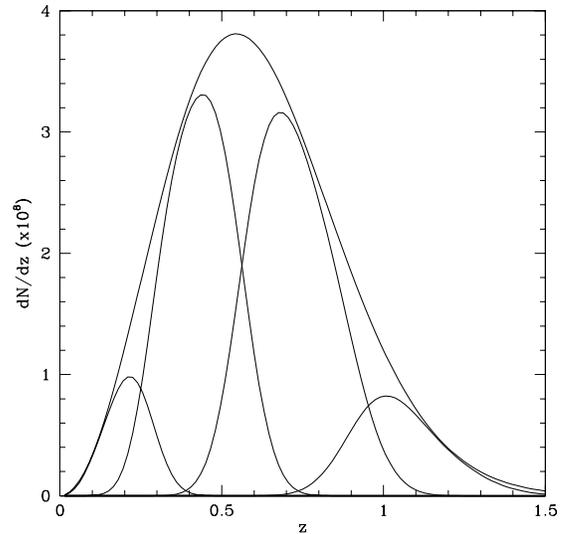}}
\caption{The galaxy number distribution {\it vs.}~redshift expected from the DES survey.}
\label{fig:des}
\end{figure}
\begin{figure}[tbp]
\centering
\scalebox{0.38}{\includegraphics{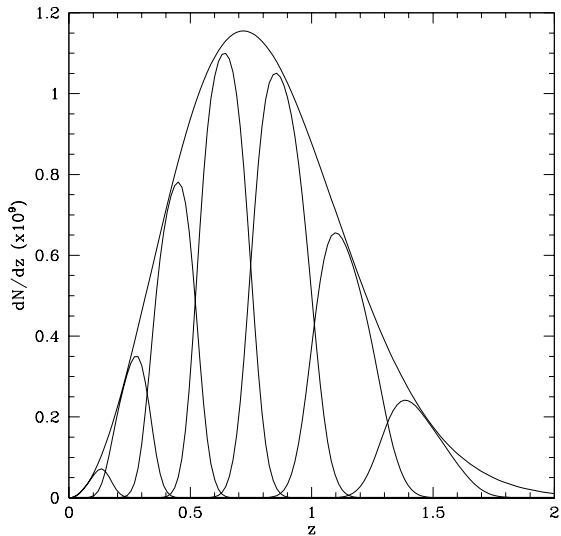}}
\caption{The galaxy number distribution {\it vs.}~redshift for the
``conservative'' LSST survey.}
\label{fig:lsst_cons}
\end{figure}
\begin{figure}[tbp]
\centering
\scalebox{0.38}{\includegraphics{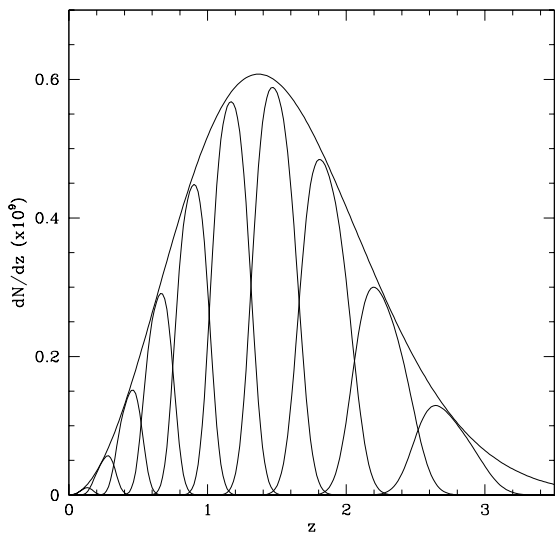}}
\caption{The galaxy number distribution {\it vs.}~redshift for the ``goal'' LSST survey.}
\label{fig:lsst_goal}
\end{figure}
To describe the distribution of galaxies from a given survey, we take the total 
galaxy number density to have the following redshift dependence \cite{hu_scran}:
\begin{equation}
n_g^{\mathrm {tot}}(z) \propto z^2 e^{-(z/z_n)^2},
\end{equation}
where $z_n$ is a parameter that is adjusted to reproduce
the expected median redshift of the survey.
The galaxies are further divided into photometric redshift bins, i.~e.
\begin{equation}
n_g^{\mathrm {tot}}(z) = \sum_i n_g^{i}(z).
\end{equation}
Following reference \cite{hu_scran}, we assume that the photometric 
redshift errors are Gaussian distributed, 
and that their rms fluctuations increase with redshift as 
$\sigma(z)=\sigma_\mathrm{max}(1+z)/(1+z_\mathrm{max})$.
The bins sizes are chosen to increase proportionally to the errors. 
The resulting photometric redshift distributions are given by,
\begin{equation}
n_g^{i}(z) = \frac{1}{2} n_g^{\mathrm{tot}}(z)\left[
  \mathrm{erfc}\biggl( \frac{z_{i-1}-z} {\sqrt{2} \sigma(z)}\biggr)
- \mathrm{erfc}\biggl( \frac{z_i-z}{\sqrt{2} \sigma(z)}\biggr) \right],
\label{eq:erfc}
\end{equation}
where erfc is the complementary error function.

For a given photometric redshift bin, the normalized selection function is given by
\begin{equation} \label{windowfunction}
 W_g^{i}(z) = \frac{n_g^{i}(z)}{\bar{N}^i} \,
\end{equation}
where $\bar{N}^i$ is the total number of galaxies in the $i$-th bin.
The Poisson noise is uncorrelated 
between bins and contributes only to the galaxy angular auto-correlation as
\begin{equation} 
 N^{(i,j)}_\ell = \frac{\delta_{ij}}{\bar{n}_A^i},
\end{equation}
where $\delta_{ij}$ is the Kronecker delta and $\bar{n}_A^i$ is the galaxy number per solid angle in the $i$-th bin. 
The observed correlation 
$\tilde{\omega}^{(i,j)}_\ell$ is the sum of the signal and the Poisson noise:
\be \label{galaxy:noise}
\tilde{\omega}^{(i,j)}_\ell \equiv \omega^{(i,j)}_\ell + N^{(i,j)}_\ell \,
\ee

The Dark Energy Survey (DES) is designed to probe the redshift range $0.1 < z < 1.3$ 
with an approximate 1-$\sigma$ error of $0.1$ in photometric redshift.
This roughly corresponds to four bins (see Fig. \ref{fig:des}) and
a median redshift of $z=0.6$. The total expected number of
galaxies is approximately $250$ million in a $5000$ sq. deg. area
on the sky, or $f_{sky}=0.13$.

The proposed LSST survey is expected to cover up to $20,000$ sq. deg. of the sky and catalogue
over a billion galaxies out to $z\sim 3$. We consider two cases: the conservative 
scenario case with $f_{sky}=0.3$ and $7$ photometric redshift bins out to $z\sim 1.5$, 
and the desired case with $f_{sky}=0.5$ and $10$ photometric redshift bins out to $z\sim 3$.
For both cases we assume $70$ gal$/$arcmin$^2$. The assumed galaxy number distributions for
the two cases are shown in Figs. \ref{fig:lsst_cons} and \ref{fig:lsst_goal}.
In what follows, we will refer to the two LSST cases as ``{\it conservative}'' and ``{\it goal}''.

The photometric redshift errors used in our analysis should be seen as the optimistic
values for the respective experiments. For a discussion of potential sources of 
systematic errors in photo-z estimates the reader is referred to \cite{Huterer05,Ma05}.

\subsection{CMB data: WMAP and Planck}

The latest expected values for the sensitivity and the resolution parameters of 
the 4-year WMAP and 14 months Planck missions are available in \cite{wmap,planck}. 
To make comparison with previous results easier, we opted to use similar, 
although somewhat different, parameters employed in \cite{Rocha,Khoury} for the three highest
frequency channels of WMAP and the lowest three Planck HFI channels.
We have checked that using the exact parameters given in \cite{wmap, planck} would
lead to an insignificant modification of our results.
The relevant parameters are listed in Table~\ref{tab:wmap}.
\begin{table}[thb]
\caption{ The relevant parameters of WMAP \cite{wmap} and Planck \cite{planck}. 
We use the three
highest frequency WMAP channels and the lowest Planck HFI frequency channels.}
\label{tab:wmap}
\begin{center}
\begin{tabular}{@{~}l@{~~~}r@{~~}r@{~~}r@{~~~~~}r@{~~}r@{~~}r@{~}} 
& \multicolumn{3}{c}{WMAP}&\multicolumn{3}{c}{Planck} \\
\hline
$\nu$ (GHz)                        & 41 & 61 & 94   & 100  & 143  & 217\\
\hline
$\theta_{\mathrm{FWHM}}$ (arc min) & 28 & 21 & 13   & 10.7 & 8.0  & 5.5\\
\hline
$\sigma_T$ ($\mu$K)             & 22 & 29 & 49 & 5.4  & 6.0  & 13.1\\
\hline
$\sigma_E$ ($\mu$K)             & 30 &  45   &  75  & n/a  & 11.4 & 26.7\\
\hline
$f_\mathrm{sky}$                &       &  0.8     &        & ~    & 0.8 & ~   \\
\hline
\end{tabular}
\end{center}
\end{table}
The noise contribution to the CMB temperature auto-correlation (TT) and the E-mode 
polarization (EE) spectra from one frequency channel is given by
\begin{eqnarray}
N_{\ell,c}^{AA} =\bigg( \frac{\sigma_{A,c} \ \theta_{\mathrm{FWHM},c} }{T_{\mathrm{CMB}}} \bigg)^2
e^{\ell(\ell+1)\theta^2_{\mathrm{FWHM},c}/8 \ln 2},
\nonumber
\end{eqnarray}
where $c$ labels the channel and $A=T,E$. The combined noise from all channels is 
\be
N_\ell^{AA}=\left[ \sum_c (N_{\ell,c}^{AA})^{-1} \right]^{-1} \ .
\ee
The observed spectra, i.e. the signal plus the noise, are then
\be
\tilde{C}^{AA}_\ell \equiv C^{AA}_\ell + N^{AA}_\ell .
\label{cmb:noise}
\ee



\subsection{SN data: SNLS and SNAP}
\begin{table*}
\caption{The redshift distribution of type Ia supernovae $N(z)$ for SNLS \cite{snls} and SNAP
\cite{kim_etal}, together with 300 SNe from the NSNF.
The redshifts given are the upper limits of each bin.
Magnitude errors $\sigma_m(z)$ are evaluated at bin midpoints.\label{tab:SN}}
\begin{center}
\begin{tabular}{|@{~}c@{~}|l@{     }|c@{   }c@{   }c@{   }c@{   }c@{   }c@{   }c@{   }c@{  }
c@{   }c@{   }c@{   }c@{   }c@{   }c@{   }c@{   }c@{   }c|@{}}
\hline
\multicolumn{1}{|c}{}&\multicolumn{1}{r|}{$z \rightarrow$} &~0.1~&~0.2~&~0.3~&~0.4~&~0.5~
&~0.6~&~0.7~&~0.8~&~0.9~&~1.0~&~1.1~&~1.2~&~1.3~&~1.4~&~1.5~&~1.6~&~1.7\\
\hline
\multirow{2}{*}{SNLS}
& $N(z)$ &~300 &~56 &~70 &~84 &~133 &~105 &~84 &~84 &~42 &7 & & & & & & &\\
& $\sigma_m(z)$ ($\times 10^{-3}$)~&~9&~20&~19&~18&~16&~19&~22&~23&~29&~60& & & & & & &\\
\hline
\multirow{2}{*}{SNAP}    
& $N(z)$ &~300~&~35~&~64~&~95~&~124~&~150~&~171~&~183~&~179~&~170~&~155~&~142~&~130~&~119~
&~107~&~94~&~80\\
& $\sigma_m(z)$ ($\times 10^{-3}$)~&~9~&~25~&~19~&~16~&~14~&~14~&~14~&~14~&~15~&~16~&~17~
&~18~&~20~&~21~&~22~&~23~&~26\\
\hline
\end{tabular}
\end{center}
\end{table*}

The DE constraints from supernovae surveys depend on the depth of the survey,
the total number of the supernovae and their redshift distribution,
i.~e. on the number of SNe per redshift bin. For example, when constraining an 
evolving equation of state, it is known \cite{frieman_etal} that the choice of the 
distribution affects the bounds on the relevant parameters (\eg~ $[w_0$,$\delta_w$]).

SNe observations determine the magnitude, $m(z)$, defined in eq.~(\ref{def:mz}).
The uncertainty in $m(z)$, at any $z$-bin containing $N_{\mathrm{bin}}$ supernovae,
is given by
\begin{equation}
\sigma_m(z) = \sqrt{\frac{\sigma_\mathrm{obs}^2}{N_{\mathrm{bin}}}+ \mathrm d m^2},
\label{sigma_m}
\end{equation}
and we assume $\sigma_\mathrm{obs} = 0.15$.
The systematic error, $\mathrm d m$, is assumed to increase linearly with redshift:
\begin{equation}
\mathrm d m = \delta m \frac{z}{z_{\mathrm{max}}},
\end{equation}
$\delta m$ being the expected uncertainty and $z_{\mathrm{max}}$ the maximum
redshift.

The SNLS is expected to measure approximately 700 supernovae out to
$z_{\mathrm{max}} = 1$ with an uncertainty $\delta m = 0.02$.
The SNLS supernovae distribution that we use for our short term forecasts
is shown in Table \ref{tab:SN}. The ground-based NSNF experiment 
\cite{nsfactory} will add to the count ~300 SNe at $z \lesssim 0.1$.

SNAP will provide over 2000 supernovae with $\delta m =  0.02$ and $z_{\mathrm{max}} = 1.7$.
In this work we have considered a fiducial SNAP survey analogous to the one
modeled in \cite{kim_etal}. The parameters are given in Table \ref{tab:SN}.

In addition to supernovae,
SNAP can get over 300 million galaxies (at $n_{gal}=100$ arcmin$^{-2}$)
at moderate and high redshifts, covering potentially up to $10,000$ sq.deg of the
sky. This would make it a candidate survey for ISW studies
\footnote{We thank Eric Linder for pointing this out}.

\section{Results and Discussion}\label{results}
\subsection {Cross correlations and $\mathbf{w(z)}$}

As discussed earlier, CMB is sensitive primarily to a particular weighted 
average of the dark energy equation of state, given by 
$\left<w\right>$ \cite{huey99,dave02,saini03}.
Measurements of the acceleration by SNe are predominantly sensitive to the very low
redshift behavior of $w(z)$.  The cross correlation measurements,
on the other hand, can give an independent window on the behavior of the 
equation of state at intermediate redshifts.
In this subsection we justify our expectation for the CMB/LSS
cross-correlation to be an effective probe of $w(z)$.
Additional analysis of the dependence of the cross
correlation on the time evolution of the DE equation of state
can be found in \cite{GPV04,Levon04}.



First note that in the small angle limit, the cross correlation can be written as
\be
X(\theta \rightarrow 0) 
\approx {const} \int dz \ W_g(z) D(z) {d \over dz}\left[(1+z)D(z)\right] \ .
\label{x-approx}
\ee
where $D(z) \equiv \delta(k,z)/\delta(k,0)$ is the linear growth factor and we
used the Poisson equation to express $\Phi$ in terms of $\delta$: 
$\Phi(k,z) \propto (1+z)\delta(k,z)$. The small angle limit $X(\theta \rightarrow 0)$ 
is representative of the total cross-correlation signal inside the redshift window 
$W_g(z)$ \cite{Levon04}. 

Eq.~(\ref{x-approx}) shows that the cross-correlation is essentially the product of 
the growth function 
and its derivative averaged over a given range of redshifts.
Having several weakly overlapping selection functions $W_g^{i}(z)$, as in the case of
LSST (\eg~see Fig.~\ref{fig:lsst_goal}), would allow one to effectively map the evolution of this 
product. Since the rate of the growth is a particularly sensitive probe of the details
of the dark energy evolution (see \cite{CHB03} for a discussion), cross-correlation studies can
provide useful information despite large statistical errors.
\begin{figure}[tbp]
\centering
\scalebox{0.38}{\includegraphics{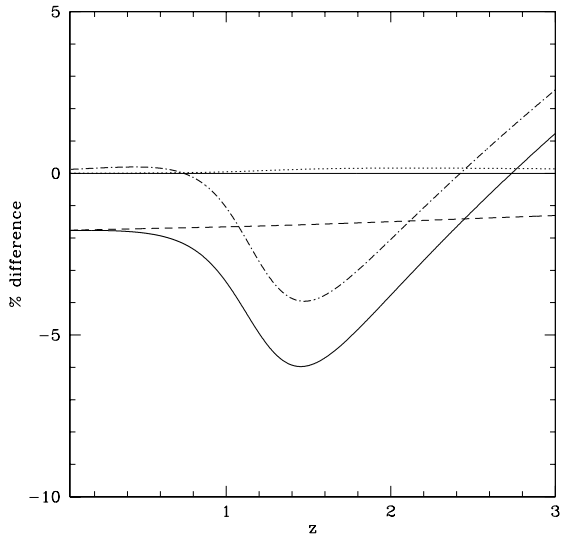}}
\caption{The percent change in $P(z)$ (eq.~(\ref{pz}), solid line), 
$d_L(z)$ (dot), $[D(z)]^2$ (dash) and $\Omega_{DE}(z)$ (dash-dot) caused by
a $10$\% change in $\Delta_w$, while holding $\left< w \right>$ fixed.}
\label{fig:10per}
\end{figure}
To illustrate the sensitivity of the cross-correlation to changes in the 
evolution of $w(z)$, we look at the change in the quantity
\be
P(z) \equiv D(z) {d \over dz}\left[(1+z)D(z)\right]
\label{pz}
\ee
caused by a $10$\% decrease in the parameter $\Delta_w$ of our Model~II. 
We hold the other two DE parameters, $\left< w \right>$ and $\Delta_z$, as well
as the cosmological parameters, fixed at their fiducial values. 
Then we compare
the resulting relative difference in $P(z)$ to the differences in $[D(z)]^2$
and in $d_L(z)$.
As one can see in Fig.~\ref{fig:10per},
$P(z)$ is by far the more sensitive probe of the three. 
Also note, that the
evolution of $P(z)$ is almost identically tracing the evolution of the 
DE fraction $\Omega_{DE}(z)$. This is expected, since the change in the gravitational
potential $\Phi(z)$ is directly caused by a non-zero $\Omega_{DE}(z)$. In particular,
it is shown in \cite{Wang98} that the evolution of $(1+z)D(z)$ couples to $w(z)\Omega_{DE}(z)$.

It is possible to explain the differences in Fig.~\ref{fig:10per} qualitatively.
The reason $d_L(z)$ is not sensitive to a change in 
$\Delta_w$ is mainly due to the change occurring at a high redshift: $z_T \sim 1.2$.
The main contribution to $d_L(z)$ in the integral given by eq.~(\ref{lumdis}) comes from 
low redshifts, where $w(z)$ is not affected by a change in $\Delta_w$. That is, $w=-1$
at low redshifts for any small variation around Model~II.

A change in $\Delta_w$ (while holding $\left< w \right>$ fixed) 
affects the observables via a combination of two effects. One is the change in 
the high redshift value of $w$, which alters the DE fraction at early times. 
The second is the change in the transition redshift, $z_T$.
The change to the growth factor comes mainly from the first of the two effects,
while $P(z)$ is affected by both.

The sensitivity of $P(z)$ to $w(z)$ is off-set by the large variance in 
cross-correlation measurements. Below we present
results of the Fisher analysis that reflect this limitation.

\subsection{Bias}
\label{sec:bias}

\begin{figure}[tbp]
\centering
\scalebox{0.38}{\includegraphics{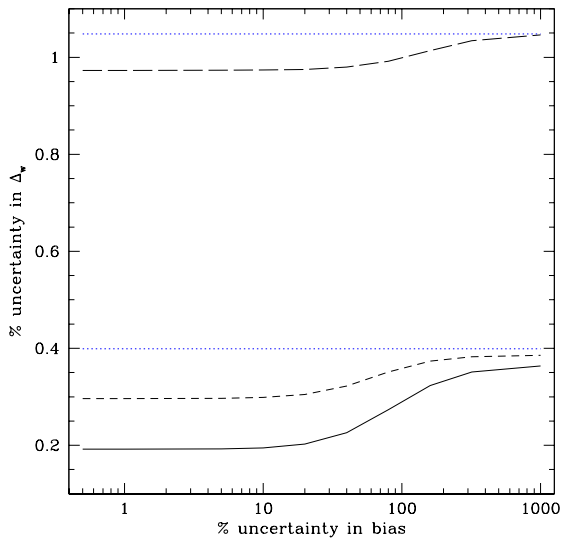}}
\caption{The dependence of the uncertainty in the Model~II parameter $\Delta_w$ on the
assumed bias prior. The long dash line corresponds to the error from the CMB/LSS cross-correlation
using WMAP and DES, the short dash-line from Planck and the ``conservative'' LSST, and the solid line
from Planck and the ``goal'' LSST. The blue dot lines are the expected errors in $\Delta_w$ from
the WMAP data alone (the upper dotted line) and from Planck alone (the lower dotted line) 
shown for reference.}
\label{fig:bias}
\end{figure}

Most of the contribution to the CMB/LSS correlation signal comes from large scales
\footnote{The actual physical scale depends on the redshift of the selection function
and on the model. For reference, for galaxies at $z\sim 0.2$ the dominant cross-correlation
scale for the model with a constant $w=-0.9$ is around $50$Mpc/h\cite{GPV04}. 
On smaller scales the ISW effect is negligible (the potential wells have to be long enough
for photons to notice the effect of stretching by the accelerated expansion). 
On very large scales the cross-correlation signal naturally decreases with the reduction in 
clustering.},
where perturbations are well described by the linear theory.
On such scales, galaxies are expected to closely trace the distribution of dark matter,
up to a scale-independent bias factor $b_g$. The bias is also known to vary with redshift,
hence, each of the photometric bins will, in principle, have a different bias factor.
The bias factors corresponding to each bin can be determined from the amplitude 
of the primordial spectrum extracted from the CMB and the galaxy-galaxy autocorrelation 
spectra. It is reasonable to assume that the bias in each bin can be determined with
$10$\% accuracy.

We account for the bias uncertainty by assigning an independent constant
bias factor to each bin and treating them as parameters in our Fisher analysis.
We assume that the value of the bias can be determined from elsewhere
to a $10$\% accuracy in all bins. This assumption, which we implement by
imposing a prior, does not have a big influence on our final results,
as long as the bias in each bin is known to better than $20$\%. 
This is illustrated in Fig.~\ref{fig:bias}, where we plot the predicted Fisher 
error in $\Delta_w$ as a function of the assumed uncertainty in bias.
The fiducial values of the bias parameters were chosen to be the same
for all bins. Choosing them according to a fixed function of redshift
does not noticeably affect the results of the Fisher analysis.

\subsection{Short term prospects}
\begin{figure}[tbp]
\centering
\scalebox{0.5}{\includegraphics{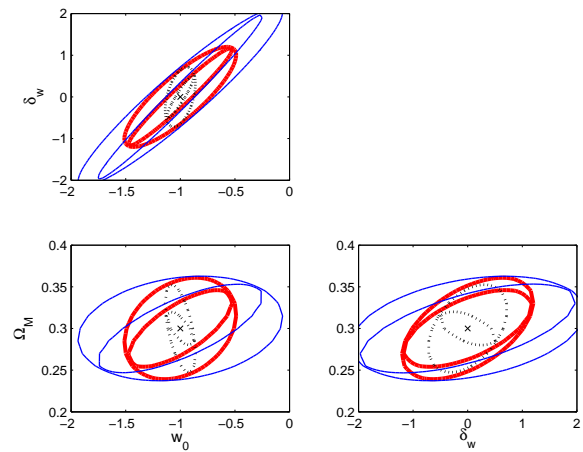}}
\caption{The expected $1\sigma$ projected contours for the $\Lambda$CDM fiducial model from WMAP (thin blue),
and the information from WMAP combined with the WMAP/DES correlation (thick solid red)
and SNLS supernovae (black dot),
using the linear parameterization of eq.~(\ref{w:linder}). For each line-type, 
the smaller ellipses correspond to
including the CMB polarization information, while the larger ellipses are obtained
using the CMB temperature spectra only. A prior from HSTKP on the value of $h$ is used.}
\label{fig:lcdm_short}
\end{figure}
\begin{figure}[tbp]
\centering
\scalebox{0.5}{\includegraphics{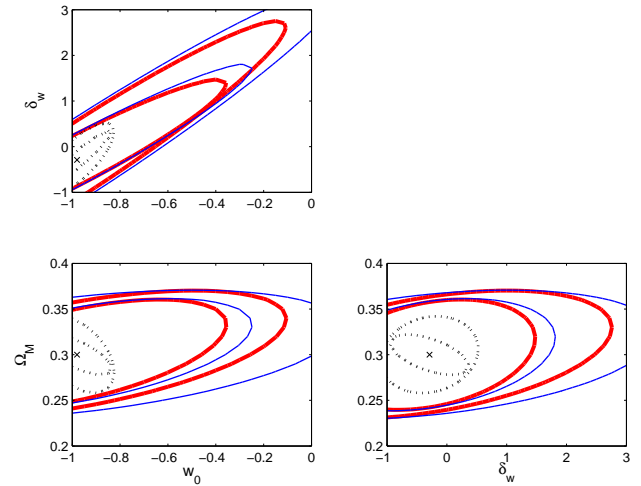}}
\caption{Short term forecasts of uncertainties on the parameters of Model~I. 
Shown are the $1\sigma$ projected contours from WMAP (thin blue), and the WMAP information combined
with the WMAP/DES correlation (thick solid red)
and SNLS supernovae (black dot). As in Fig.~\ref{fig:lcdm_short}, 
the smaller ellipses correspond to including the CMB polarization information, 
while the larger ellipses are obtained using the CMB temperature spectra only.
A prior from HSTKP on the value of $h$ is used.}
\label{fig:comb1_short}
\end{figure}
In the next few years we will have results from DES, SNLS and the complete 
data from WMAP, which will include E-mode polarization. In this subsection
we show and compare the expected constraints on $w(z)$ from these measurements.

The linear parameterization 
of $w(z)$ (eq.~(\ref{w:linder})) with fiducial parameters corresponding to $\Lambda$CDM
($w_0=-1$, $\delta_w=0$) is the most commonly considered case in the existing literature.
For that reason we include it in our analysis and show the short term prediction for 
the DE parameters in Fig.~\ref{fig:lcdm_short}. Shown are only the contours for
$\Omega_M$, $w_0$ and $\delta_w$, with other parameters being marginalized over.
In this case, the cross-correlation information provides only a modest improvement
over the CMB spectra alone. Supernovae data, on the other hand, can tighten
the allowed range of the DE model parameters somewhat better and significantly 
constrain $w_0$.
\begin{table*}[hbt]
\hspace{-1cm}
\caption{Short term predictions assuming WMAP for CMB, SNLS+NSNF for SNe and WMAP/DES 
for the CMB/LSS cross-correlation ($X$). The SNe and X results are shown with priors 
from CMB temperature only (T) and CMB temperature and polarization data combined (TP).
A prior from HSTKP on the value of $h$ is used.}
 \label{tab:short}
\begin{tabular}{|ll|c@{~}c|c@{~~}c|c@{~~}c|c@{~}c|c@{~~}c|c@{~~}c|}
\cline{1-14}
\multicolumn{2}{|c|}{}& \multicolumn{6}{c|}{Model~I, $1\sigma$ errors}&
\multicolumn{6}{c|}{Model~II, $1\sigma$ errors} \\
\cline{3-14}
 & & \multicolumn{2}{c|}{CMB}& \multicolumn{2}{c|}{SNe}&
\multicolumn{2}{c|}{X }
   & \multicolumn{2}{c|}{CMB}& \multicolumn{2}{c|}{SNe}&
\multicolumn{2}{c|}{X }\\
p                  &${\rm fiducial} \atop {\rm values}$
                          & T    & TP   &   T  &  TP  &  T   &  TP  & T    & TP  &  T  &  TP &  T   &  TP   \\ \hline
$w_0$              & -0.98   & 1.1  & 0.73 & 0.16 &0.15  &0.87  & 0.62 & -    & -   & -   & -   & -    & -     \\
$\delta_w $        & -0.29 & 3.8  & 2.1  & 0.82 &0.66  &3.0   & 1.8  & -    & -   & -   & -   & -    & -     \\
$\left< w \right>$ & -0.9 & -    & -    &   -  &  -   &-     & -    & 0.26 &0.19 &0.22 &0.13 & 0.23 & 0.17  \\
$\Delta_w$         & -0.6 & -    & -    &   -  &  -   &-     & -    & 3.2  &1.0  &2.9  &1.0  & 2.0  & 0.97  \\
$\Delta_z$         & 0.3  & -    & -    &   -  &  -   &-     & -    & 8.9  &6.2  &4.0  &2.6  & 6.6  & 4.7   \\
$\Omega_M$         & 0.3  & 0.07 & 0.06 & 0.04 &0.02  &0.07  & 0.06 & 0.07 &0.06 &0.05 &0.03 & 0.07 & 0.05 \\
$h$                & 0.69 & 0.07 & 0.07 & 0.03 &0.02  &0.07  & 0.07 & 0.08 &0.07 &0.06 &0.04 & 0.07 & 0.06  \\
$10^3\omega_b$      & 24  & 2    & 0.7  & 2    &0.6   &2     & 0.7  & 3.4  &0.8  &3.2  &0.7  & 2.7  & 0.7 \\
$n_s$              & 0.99 & 0.08 & 0.02 & 0.07 &0.02  &0.07  & 0.02 & 0.13 &0.02 &0.12 &0.02 & 0.1  & 0.02  \\
$\tau$             & 0.166& 0.17 & 0.02 & 0.15 &0.02  &0.15  & 0.02 & 0.22 &0.02 &0.22 &0.02 & 0.2  & 0.02   \\
$A_s$              & 0.9  & 0.3  & 0.04 & 0.28 &0.04  &0.27  &0.04  & 0.4  &0.04 &0.39 &0.04 & 0.35 & 0.04   \\
 \hline
 \end{tabular}
\end{table*}

The short term predictions for Model~I are shown 
in Fig.~\ref{fig:comb1_short}, again, for $\Omega_M$, $w_0$ and $\delta_w$ only.
The list of predicted uncertainties in all parameters of Model~I is given 
in Table~\ref{tab:short}. The expected uncertainties are different, and generally
larger than in the $\Lambda$CDM fiducial case. This illustrates the importance of
the underlying DE model for constraining $w(z)$.
The cross-correlation will not, in short term, provide competitive constraints
on the evolution of $w(z)$ according to $\Lambda$CDM or Model~I. 

\begin{figure}[tbp]
\centering
\scalebox{0.5}{\includegraphics{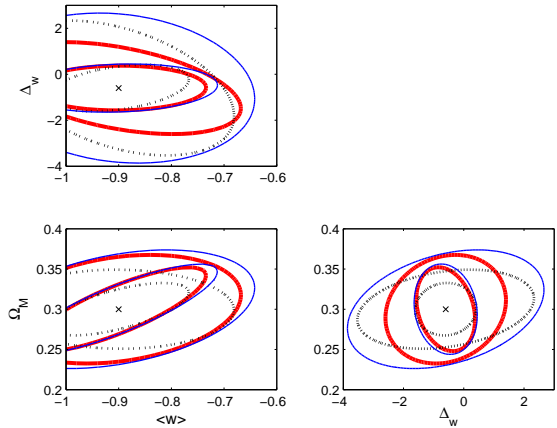}}
\caption{The expected $1\sigma$ projected contours from WMAP (thin blue), and the WMAP information
combined with
the WMAP/DES correlation (thick solid red)
and SNLS supernovae (black dot) for Model~II.
The smaller ellipses correspond to
including the CMB polarization information, while the larger ellipses are obtained
from the CMB temperature spectra alone. A prior from HSTKP on the value of $h$ is used.}
\label{fig:comb2_short}
\end{figure}

The forecasts for Model~II parameters $\left< w \right>$, $\Delta_w$ and $\Omega_M$ 
are shown in Fig.~\ref{fig:comb2_short}, while Table~\ref{tab:short} contains 
the full list of the expected parameter uncertainties. 
The choice of the
model clearly makes a big difference. Most notably, neither the short term SNe nor the 
cross-correlation measurements can improve the CMB bounds on $\Delta_w$.
The SNe, on the other hand,
are still useful in tightening the constraint on $\Omega_M$ and $\left< w \right>$.
Overall, it is clear from the plots and the table that measurements of the
cross-correlation in the next five years will not provide any new competitive 
constraint on the evolution of $w$.
In the case of Model~II this is primarily due to limited depth of the DES survey,
which does not
sample the redshifts above $z\sim 1$, where the transition in $w(z)$ occurs.
In the case of Model~I, constraints from the cross-correlation are even weaker, even though it 
has one less parameter as compared to Model~II.
It is worth noticing that for this model, 
the degeneracy between $w_0$ and $\delta_w$ that hinders their determination from CMB
alone (with a prior on $h$) is slightly improved by 
the addition of the cross-correlation information.
This is because the angular diameter distance
to last scattering surface depends on the dark energy through the average equation of
state value $\left< w \right>$. Therefore in Model I
different values of $w_0$ and $\delta_w$ 
can give the same $\left< w \right>$. On the other hand
redshift dependent measurements of the ISW-correlation are sensitive to both
the average equation of state and its time evolution improving the constraints
on these two parameters.

In summary, the short term cross-correlation measurements will not be competitive, while
supernovae measurements will only constrain $w_0$ of Model~I and $\left< w \right>$ of Model~II.
As we will see in the next subsection, the constraints on all parameters tighten in 
the case of deeper surveys such as LSST.

\subsection{Long term prospects}
\begin{figure}[tbp]
\centering
\scalebox{0.55}{\includegraphics{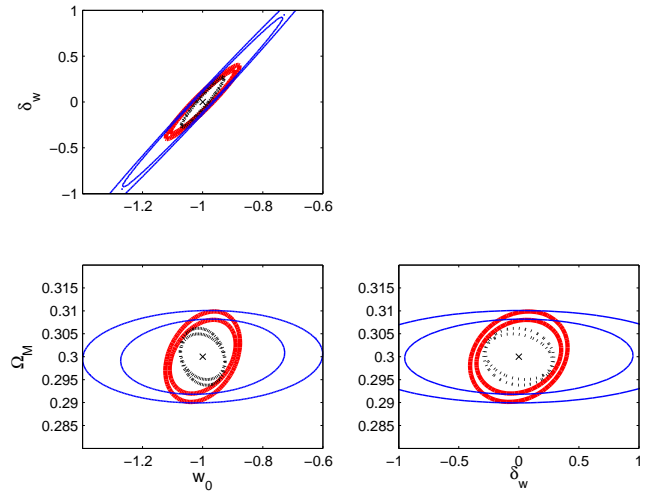}}
\caption{
The expected $1\sigma$ projected contours from Planck (thin blue), and Planck information combined with
the Planck/LSST(goal) correlation (thick solid red)
and SNAP supernovae (black dot) for the linear parameterization of 
eq.~(\ref{w:linder}) with $\Lambda$CDM fiducial parameters.
(Compare this to Fig.~\ref{fig:lcdm_short}).}
\label{fig:lcdm_long}
\end{figure}
\begin{figure}[tbp]
\centering
\scalebox{0.5}{\includegraphics{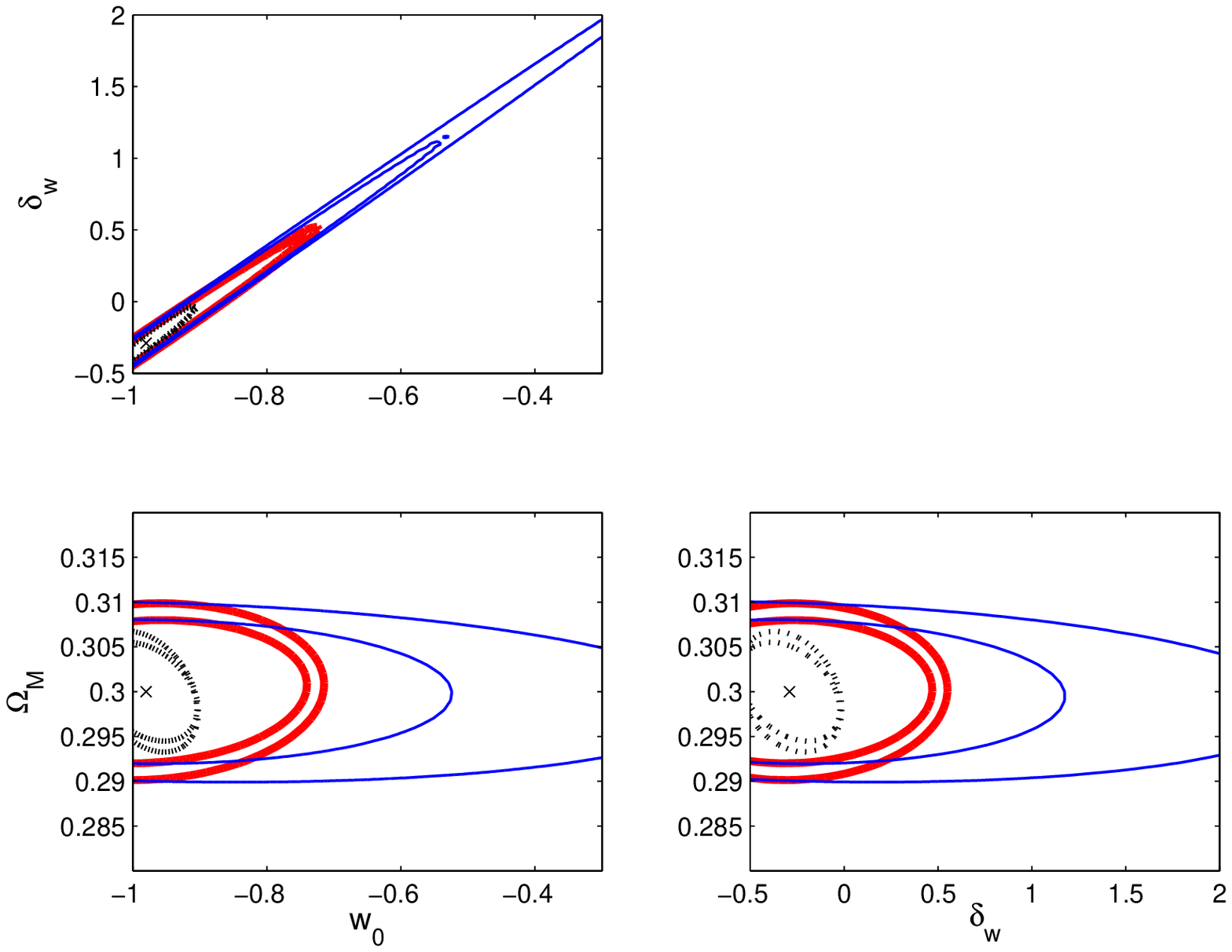}}
\caption{The expected $1\sigma$ projected contours for the Model~I parameters from Planck (thin blue), 
and Planck information combined with the Planck/LSST(goal) correlation (thick solid red)
and SNAP supernovae (black dot). (Compare this to Fig.~\ref{fig:comb1_short}).}
\label{fig:comb1_long}
\end{figure}

\begin{figure}[tbp]
\centering
\scalebox{0.55}{\includegraphics{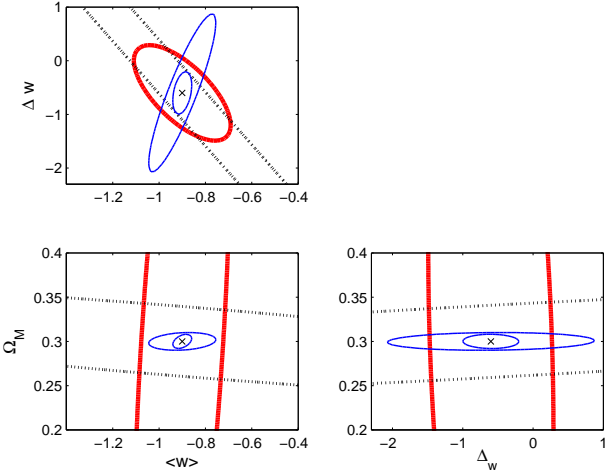}}
\caption{The expected $1\sigma$ projected contours for the Model~II parameters from the three probes: Planck 
(thin blue), Planck/LSST(goal) correlation (thick solid red) and SNAP supernovae (black dot).}
\label{fig:m2_ind_long}
\end{figure}

\begin{figure}[tbp]
\centering
\scalebox{0.55}{\includegraphics{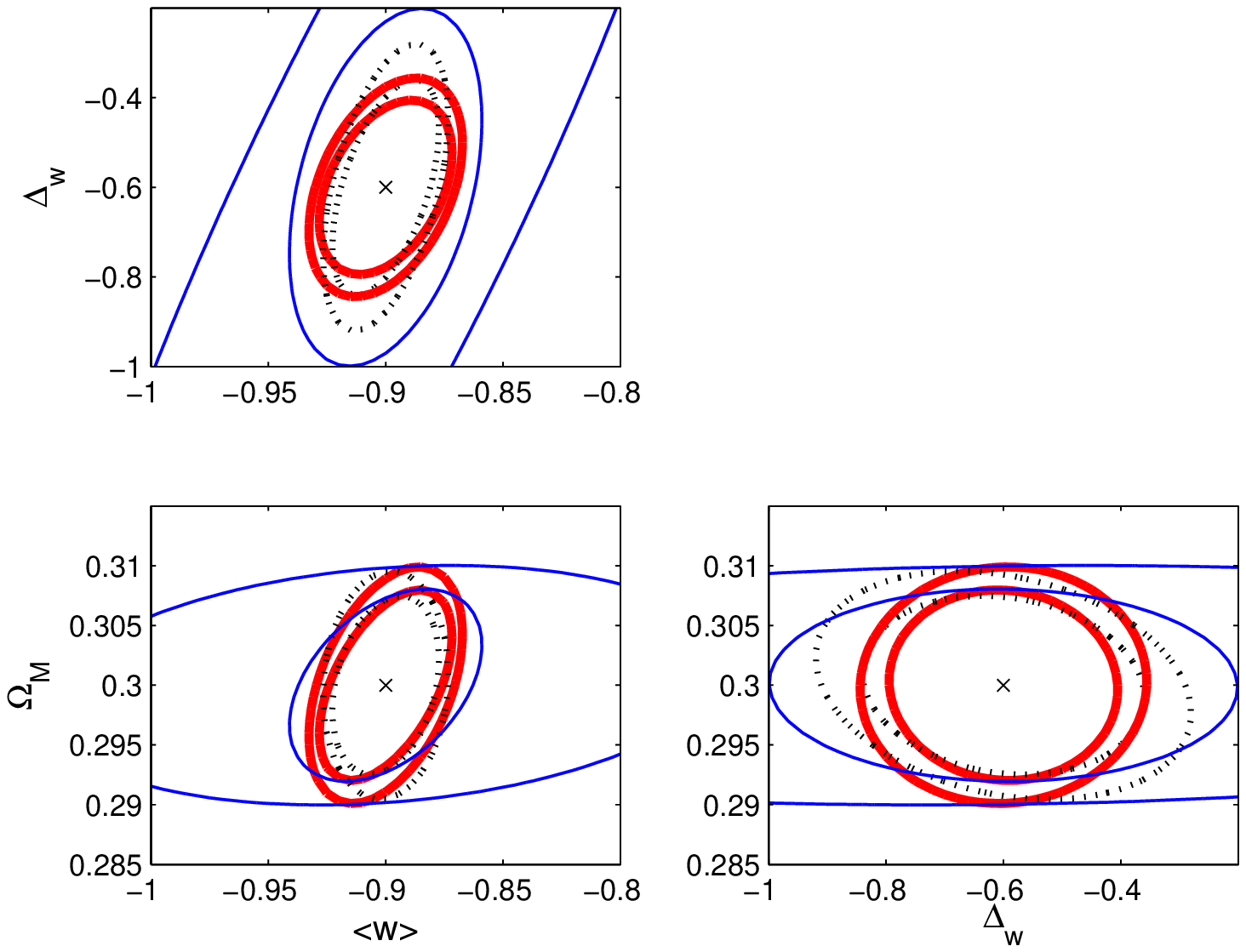}}
\caption{
The expected $1\sigma$ projected contours for the Model~II parameters from Planck (thin blue),
and Planck information combined with the Planck/LSST(goal) correlation (thick solid red)
and SNAP supernovae (black dot). (Compare this to Fig.~\ref{fig:comb2_short}).}
\label{fig:comb2_long}
\end{figure}

\begin{figure}[tbp]
\centering
\scalebox{0.55}{\includegraphics{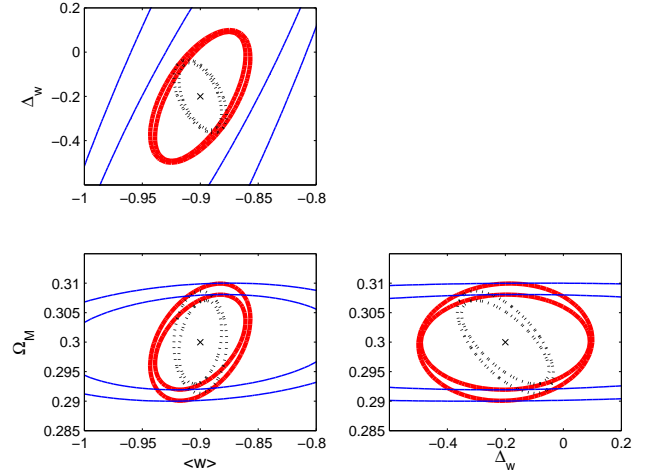}}
\caption{Same as Fig. \ref{fig:comb2_long} but using a fiducial $\Delta_w = -0.2$.}
\label{fig:alt_comb2_long}
\end{figure}

\begin{table*}[hbt]
\caption{Long term predictions assuming Planck for CMB, SNAP+NSNF for SNe and Planck/LSST 
for the CMB/LSS cross-correlation ($X$) for the``conservative'' and ``goal'' cases. 
The SNe and $X$ results are shown with priors from CMB temperature only (T) and CMB temperature 
and polarization data combined (TP).}
 \label{tab:long}
\hspace{-1cm}
\begin{tabular}{|ll|c@{~}c|c@{~~}c|c@{~~}c|c@{~}c|c@{~~}c|c@{~~}c|c@{~~}c|c@{~~}c|}
\cline{1-18}
\multicolumn{2}{|c|}{}& \multicolumn{8}{c|}{Model~I, $1\sigma$ errors}&
\multicolumn{8}{c|}{Model~II, $1\sigma$ errors} \\
\cline{3-18}
 & & \multicolumn{2}{c|}{CMB}& \multicolumn{2}{c|}{SNe}&
\multicolumn{2}{c|}{X(conserv.) } & \multicolumn{2}{c|}{X(goal) }
   & \multicolumn{2}{c|}{CMB}& \multicolumn{2}{c|}{SNe}&
\multicolumn{2}{c|}{X(conserv.) } & \multicolumn{2}{c|}{X(goal) }\\
   p              &${\rm fiducial} \atop {\rm values}$
                          & T    & TP   &  T   &  TP  &  T   &  TP  &  T  &  TP & T     & TP   &  T   &  TP  &  T    &  TP   &  T    &  TP   \\
\hline
$w_0$              & -0.98   & 0.86  & 0.46 & 0.08&0.08  &0.41  & 0.33 &0.26 & 0.24& -     & -    & -   & -    & -      & -     & -     & -     \\
$\delta_w $        & -0.29 & 2.8  & 1.5  & 0.28 &0.27  &1.3   & 1.0  &0.84 & 0.76& -     & -    & -   & -    & -      & -     & -     & -     \\
$\left< w \right>$ & -0.9 & -    & -    &  -   &  -   & -    & -    & -   & -   &0.14   &0.04  &0.03  &0.02  &0.04   & 0.03  &0.03   & 0.03  \\
$\Delta_w$         & -0.6 & -    & -    &  -   &  -   & -    & -    & -   & -   &1.5    &0.39  &0.32  &0.24  &0.44    & 0.3  &0.24   & 0.19  \\
$\Delta_z$         & 0.3  & -    & -    &  -   &  -   & -    & -    & -   & -   &8.3    &4.5   &0.9   &0.79  &2.4    & 2.2   &1.3    & 1.2  \\
$\Omega_M$         & 0.3  & 0.01 & 0.008&0.007 &0.005 &0.01  &0.008 &0.01 &0.008&0.01   &0.008 &0.009 &0.007 &0.01   & 0.008 &0.01   & 0.008 \\
$h$                & 0.69 & 0.01 & 0.009&0.006 &0.005 &0.01  &0.009 &0.01 &0.009&0.01   &0.009 &0.01  &0.008 &0.01  & 0.009  &0.01   & 0.009  \\ 
$10^3\omega_b$     &24    &0.21  &0.16  &0.2   &0.16  &0.21  &0.16  &0.21 &0.16 &0.22   &0.16   &0.2   &0.16  &0.21  & 0.16  &0.21   & 0.16 \\
$n_s$              & 0.99 & 0.005&0.004 &0.005 &0.004 &0.005 &0.004 &0.005&0.004&0.005  &0.004 &0.005 &0.004 &0.005  & 0.004 &0.005  & 0.004 \\
$\tau$             & 0.166& 0.04 &0.006 &0.03  &0.005 & 0.03 &0.006 & 0.03&0.006&0.05   &0.007 &0.03  &0.006 &0.03   & 0.006 &0.03   & 0.006 \\
$A_s$              & 0.9  & 0.07 &0.01  &0.06  &0.01  & 0.06 &0.01  & 0.06&0.01 &0.1    &0.01  &0.06  &0.01  &0.06   & 0.01  &0.06   & 0.01  \\
 \hline
 \end{tabular}
\end{table*}

In the longer term, i.e. in the next ten years, we should expect to have the results
from Planck CMB measurements, galaxy catalogues from LSST and the SNe measurements from
SNAP. These will significantly improve the constraints on the DE parameters. In certain
models, as we show here on the example of our Model~II, CMB/LSS cross-correlation can
play a competitive role.

As in the previous subsection, we begin by considering the linear DE model with two different 
sets of the fiducial values: one where they match a cosmological constant ($w_0 = -1, \delta_w = 0$) 
and another where $w$ evolves but still consistent with present observations (Model~I).  
In Fig.~\ref{fig:lcdm_long} and Fig.~\ref{fig:comb1_long} we show the $1\sigma$ projected contours 
for the linear models. On these plots, the cross-correlation and the SNe prediction includes
the information from Planck. Combining with Planck is equivalent to applying
strong priors on all cosmological parameters. The constraints in the absence of any priors can be
found in Table~\ref{tab:long}. Planck and SNAP are by far the most dominant information
sources for these models, with cross-correlation adding a relatively weak contribution.

For Model II we start by presenting the individual error contours for the three experiments
(Fig. \ref{fig:m2_ind_long}), with the purpose of explicitely showing degeneracy directions.
We will see that, once we consider its combination with Planck, the

relative utility of the cross-correlation is significantly higher for Model~II.
The contours in Fig.~\ref{fig:comb2_long} show that the cross-correlation can 
provide constraints on $\left< w \right>$ and $\Delta_w$ that are competitive
with those from SNAP. The improvement on Planck is especially strong without the
polarization data. The full list of parameter constraints is given in Table~\ref{tab:long}.
Our results confirm that SNe measurements are more suitable for constraining variations
in $w(z)$ at low redshifts. At higher redshifts, the cross-correlation becomes useful
and can provide an independent constraint $\Delta_w$.

Figure \ref{fig:alt_comb2_long} illustrates how a different choice of Model II fiducial 
parameters affects the constraining capabilities of the different probes.
The alternative model has $\Delta_w=-0.2$ instead of $-0.6$ 
\footnote{The choice of the fiducial value of $\left< w \right>$ was not modified as the transition time would be
pushed to very high redshifts if it approached -1. On the other hand, $\left< w \right>$ smaller than $-0.9$ are strongly disfavored by current data.}.
In this case we note the SNAP+Planck constraint for $\Delta_w$ is noticeably tighter 
than that from the Planck/LSST correlation. Also note that this is mainly due to improvement
in the sensitivity of the SNe, as the transition in this model happens at a lower redshfit.
The size of the error on $\Delta_w$ from the cross-correlation is only marginally increased
compared to Fig.~\ref{fig:comb2_long}. 


\section{Conclusions}\label{conc}

We have studied the potential of the CMB/LSS correlation for constraining the
evolution of the DE equation of state. We proposed a new parameterization, a ``Kink'' model
which has the average equation of state $\left< w \right>$ as an explicit
parameter. A direct control on the value of $\left< w \right>$ preserves
the tight constraints on $w$ obtained from the distance to last scattering measurements,
while allowing $w(z)$ to vary in other ways. 

Using Fisher analysis, we have made forecasts of the expected uncertainties in DE
parameters extracted from the upcoming CMB, SNe and CMB/LSS correlation data. We have
considered both short- (WMAP, DES, SNLS) and long- (Planck, LSST, SNAP) term prospects.
We find that in long term the cross-correlation will provide competitive constraints
on the evolution of $w(z)$, in particular, on the total change in $w$,
provided a sufficiently generic parameterization is used. 
We also note that the transition length parameter, $\Delta_z$, will not be
well constrained even with the long-term combined data.

Our results encourage further investigation of the CMB/LSS correlation as a probe of
dark energy. The cross-correlation has a clear weakness -- a large variance that
hinders its utility as a precision cosmology probe. However, it also has certain
attractive features, that in some cases can out weigh the lack of accuracy. These
features are:
\begin{itemize}
\item The cross-correlation probes {\it the rate} of the growth of structure 
as well as the growth.
\item It is practically independent of reionization details ($\tau$) and
the details of the initial spectrum ($n_s$).
\item It probes the large scales ($0.01<k<0.1 h {\rm Mpc}^{-1}$), 
where the evolution of structure is safely
inside the linear regime and there is no need for higher order corrections.
\item On linear scales contributing to the ISW/LSS correlation 
the galaxy bias is expected to be independent of scale.
\item Cross-correlation is linear in bias, compared to the matter
power spectrum that is quadratic. In fact, if one used a cross-correlation 
estimator that is normalized to the auto-correlations, that estimator would
be independent of bias and the overall normalization of the primordial
spectrum.

\item Cross-correlation can probe the evolution of dark energy at relatively
high redshifts, inaccessible by luminosity distance measurements.
\end{itemize}

\

In this work we have restricted our attention to Quintessence models for which 
the effective sound speed is of order of unity, so that the dark energy 
perturbations only affect the evolution of the largest 
structures. However, if the dark energy clusters on smaller scales, it may leave a 
distinctive signature which can be detected with cross-correlation 
measurements.
Although current data do not provide any constraints on the dark energy 
sound speed
\cite{bean03,CGM}, the constraints from the next generation of galaxy surveys have 
been studied in \cite{hu_scran}. 

We have only considered flat FRW models.
Without the flatness assumption, CMB constraints on
$\Omega_M$ and $h$ weaken because of a well-known degeneracy \cite{BondEfstathiou}, 
also relaxing constraints on $w(z)$. We will explore constraints on DE using 
ISW without the flatness prior in a future work.

The Model II type parameterization of $w(z)$ considered in the
paper is not necessarily the most optimal for underscoring the
power of cross-correlation constraints (although clearly better suited for this
purposed than Model I). A more comprehensive study using a principal
component approach \cite{HutererStarkman,Crittenden05} is a subject of 
ongoing research.

Another potential use of the ISW 
cross-correlation is to measure the amplitude of the scalar perturbations 
and indirectly constrain the amplitude of the tensor modes when combined 
with the normalization of CMB spectra \cite{CCGM}. This is because the 
distribution of the large scale structures is not correlated with a 
stochastic background of gravitational waves. 
It has been also suggested \cite{glenn} that CMB/LSS studies may help to 
differentiate between Quintessence and models with modified gravity.

For a fixed
total number of galaxies the information extracted from the ISW studies
depends on the choice of the redshift binning. The role of cross-correlation in 
constraining $w(z)$ may be further
increased by optimizing the bin selection. 
Such an optimization is a topic of ongoing research.

\acknowledgments
We thank Niayesh Afshordi, Bruce Bassett, Wayne Hu, Justin Khoury and Ryan Scranton
for useful comments and discussions. We thank Eric Linder for very helpful
critical comments on the first version of the paper.
P.S.C. thanks Tufts Institute of Cosmology for hospitality.
The code used for calculating the cross-correlation was
developed in collaboration with Jaume Garriga and Tanmay Vachaspati
and was based on CMBFAST \cite{cmbfast}. P.S.C. is supported by 
Columbia Academic Quality Fund.

\end{document}